# An Enhanced Search Technique for Managing Partial Coverage and Free Riding in P2P Networks

Sabu M. Thampi[1], Chandra Sekaran K[2]
[1]L.B.S Institute of Technology for Women, Kerala-695012, India
[2]National Institute of Technology Karnataka, Surathkal, Karnataka-575025, India
smtlbs@yahoo.co.in, kch@nitk.ac.in

***Abstract*— This paper presents a Q-learning based scheme for managing the partial coverage problem and the ill effects of free riding in unstructured P2P networks. Based on various parameter values collected during query routing, reward for the actions are computed and these rewards are used for updating the corresponding Q-values of peers. Thus, the routing is done through only nodes, which have shown high performance in the past. Simulation experiments are conducted in several times and the results are plotted. Results show that the proposed scheme effectively manages free riders, generates high hit ratio, reduces network traffic and manages partial coverage problem.**

***Index Terms*—unstructured P2P, searching, replication, free riding, partial coverage**

## I. Introduction

The ineffective flooding based blind search techniques rationalize the scalability of the unstructured peer-to-peer (P2P) systems. To evade huge amount of redundant traffic uphold by flooding-based search methods, several efforts have been made to brighten the performance of search schemes. Most of these techniques fall into a category of searching called informed search schemes. Informed techniques make use of their indices to accomplish similar quality results, and to minimise overhead. The Q-learning based resource discovery scheme proposed in this paper eludes probabilistic query routing and provides two-way load-balancing, alleviates partial coverage problem, and to some extent controls free riders in the network. The informed search technique achieves good response time, high success rate, low network traffic and adaptive behaviour.

This paper is a revised version of [1]. It presents a distributed search technique (DST) for unstructured P2P networks. In the previous version the effect of free riders is not directly addressed. The free riders create excess amount of network traffic and don't share popular files in their shared files folder. Routing queries through free riders are curtailed to decrease redundant messages. Not all the peers may be reachable during searching and this creates the so-called partial coverage problem. In order to alleviate this problem the queries are routed appropriately. Moreover, the parameters and conditions for computing rewards in different situations are modified. The major operations of DST are illustrated using an example. The simulation is conducted in a different environment and more performance metrics are employed.

The remainder of this paper is organized as follows. Section II explains the model of the system. The Q-table update process is discussed in section III. An illustration of DST is given in section IV. Section V describes the experimental setup for conducting simulations. Section VI discusses the results of experiments conducted. Finally, section VII concludes the paper.

## II. Model of the System

The peer-to-peer system model comprises nodes and files. The term files represent any generic content while a node corresponds to a peer. A file can have more than one replica in the system. The number of neighbours connected (links) to a node is called its degree. The topology of P2P networks is modelled as a network with an undirected graph G whose nodes represent hosts and edges represent internet connections between those hosts. Nodes are usually very dynamic, where some can join and leave the network in the order of seconds whereas other nodes stay for an unlimited period of time. When a user requests a file, a search for the file is initiated and other nodes in the network need to be queried if the file is not available locally. A query consists of a list of desired words.

The nodes in the network are classified as ordinary nodes and power peers. In each node, the proposed distributed search technique (DST) maintains different kinds of indexes on other nodes. If the object is not found in the local directory, the node searches for matches in its index. Nodes keep information on processed queries in a table. This is the first index being searched if the object is not found in the local repository. If the query keyword exists in the table, query is further routed appropriately through walkers based on past search results. In the case the query keyword is a new one, query is routed simultaneously through walkers selected from neighbours and power peers based on index values maintained for neighbouring nodes and power peers. Usually the power peers receive more requests in contrast to ordinary nodes in the neighbour list. Hence, queries are routed suitably for balancing query load. DST forwards the duplicate messages to other peers according to the class of the peer from which the message is received and the ranks of remaining neighbours who have not yet forwarded the query. The search is terminated either result is found or time-to-live (TTL) is expired. However, the powerful peers can extend search to other powerful nodes in the network by extending the TTL value.

The index values are initialised and maintained by employing a type of reinforcement learning called Q-learning [2]. The indexes are maintained in different data structures known as Q-tables. Q-tables consist of Q-values, which are integer numbers pointing the past performance of different nodes in processing queries. Each node in a table has a corresponding Q-value. Like in search schemes such as random walk [3] and adaptive probabilistic search (APS) [4], in DST, walkers are not selected either randomly or probabilistically. Walkers are selected based on their past performance, which is depicted in Q-tables in the form of Q-values. Thus, the application of Q-learning evades probabilistic or random routing of queries to achieve improved success rate. Each search operation updates the Q-values in different Q-tables. Three kinds of update operations are associated with Q-tables. The Q-values of a node in a table is modified according to its search results. Even if the performance of one node degrades, another node in the network is suitably chosen as target node for forwarding the queries. Hence, the search technique adapts to the changing environments. The discounted future rewards are not considered as the Q-values of all nodes above a peer P in the search path are required to compute the Q-value. This flow of Q-values increase the size of the messages causing increased network traffic.

A node generally maintains three Q-tables: *query Q-table, neighbour Q-table and power peer Q-table*. The Query Q-table in a node contains the past processed query keywords along with Q-values of all neighbours. For a successful search through a neighbour, corresponding Q-value is modified according to number of hops and results returned; otherwise, penalty is awarded. For all neighbours who have responded with successful results, associated entries in the table are modified. The table grows as the entries for successful queries with new keywords are added. First time a query keyword enters, the related Q-values for all neighbours are initialised with the same Q-values available in the neighbour Q-table and these values change according to result of every search operation.

To avoid free riders in becoming neighbours, a node $n_1$ selects a node $n_2$ as neighbour if the number of objects being hosted ($f_n$) by n2 is greater than or equal to a threshold $f_{th}$. The presence of large number of objects in the neighbouring peers also increases the chance of finding out required objects quickly. However, assigning very high threshold value increases the reliance on only a small number of nodes with large quantity of objects as neighbours for all the nodes. The load on these nodes increases heavily, thus causing inferior system performance. Consequently, the value of $f_{th}$ should always be not higher than the number of objects in $n_1$. Both categories of nodes, ordinary nodes and power peers, have eligibility to become neighbours of a node. However, while powers peers are superior to other nodes in the network, they can directly become the neighbour of the node without undergoing the selection process.

Neighbour Q-table (NQ) of a node contains Q-values of neighbours of a node. When a user queries for keywords not existing in query Q-table, the query is routed through both neighbours and power peers simultaneously till TTL expires or the result is found. Hence, walkers are selected from both neighbour Q-table and power peer Q-table in the descending order of Q-values. The neighbour Q-table provides an overall picture with respect to the performance of neighbours in the past. The Q-values of neighbours are initialised as $Q_i = round\left[\left(\frac{f_i}{f_{th}}\right) * k\right]$; where the value of 'k' is preset as 100, in order to maintain a minimum of 100 as Q-value. However, the update of Q-value of a node in a neighbour Q-table relies on the success rate ($h_r$). The success rate (hit rate) is the ratio of number of hits produced and total number of queries received by a node and the success rate shows how a node responds for various search queries.

In order to extend search process through the neighbours and power peers simultaneously, each node maintains a list of nearby power peers in its power peer Q-table. It contains the address of power peers and their corresponding Q-values. The Q-values are associated with number of hops being visited during different search operations. An entry for a power peer is added to the table each time a node receives a broadcast message from a power peer or the requested object is found in another power peer, which is not listed in the power peer table. Each node, irrespective of its class it belongs as a power node or an ordinary node, maintains a variable to store the number of hits occurred in the node.

The moment a node becomes a power peer, it broadcasts a message to all the nodes within N hops away by dispatching mobile agents. Clones of mobile agents are created to visit several sites. The broadcast message is also propagated through, its neighbours as well as the power peers identified by the node. Before sending the broadcast message from a newly upgraded peer as power peer to nearby peers, the peer computes a Q-value ($Q_p$) to be used by other peers for initialisation. This Q-value is computed using the available storage ($s_s$), degree of the node ($d_d$) and number files being hosted ($f_i$) in the new power peer. Assume the threshold value for minimum storage is '$s_{th}$', degree threshold '$d_{th}$' and minimum number of files to be hosted is '$f_{th}$'. The Q-value is computed as,

$$Q_p = round\left[\left(w_1 * \left[\frac{s_s}{s_{th}}\right] + w_2 * \left[\frac{d_d}{d_{th}}\right] + w_3 * \left[\frac{f_i}{f_{th}}\right]\right) * k\right]$$

where $w_1$, $w_2$ and $w_3$ are the assigned weights, $w_1 + w_2 + w_3 = 1$ and 'k' is a constant for normalizing the Q-value and its value is always 100. The Q-value ($Q_p$) is also included in the broadcast message. If the same entry of the peer does not exist in its neighbour Q-table of the node, which has received the broadcast message, it inserts a new entry for the power peer in its power peer Q-table with its initial Q-value as $Q_p$.

During search, if the required object is found in a power peer and the entry of that peer is not listed in the power peer Q-table, it is also added to the table. After every successful search operation, the node that holds the object transmits the reply message along the reverse path. Some of the parameters in the reply message include query source-id, message-id, address of object node, and its status (power peer or ordinary peer). However, if the requested object is found in a power peer; the reply message from the peer contains one more field to include $Q_p$. In case, the node status is 'power peer', and the entry for that node is not there in the power peer Q-table, new entry for the power peer is added to the power peer Q-table by the intermediate nodes on the reply path with the initial Q-value as $Q_p$. DST impedes power peers already listed in neighbour Q-table to turn into members of power peer Q-table. The transfer of query messages simultaneously through neighbours and power peers reduce query load on power peers and at the same time, it leads to the exploration of rare objects owned by the ordinary nodes.

DST does not use random or probabilistic selection of peers for routing queries, but it employs a learning mechanism. If the query keyword exists in the query Q-table, the query is routed through the neighbours of a node based on the numbers of walkers to be used. Otherwise, both the neighbours in the neighbour Q-table as well as power peers in the power peer Q-table of the query source are chosen to route query. The selection is based on Q-values of peers in the neighbour Q-table (NQ) and power peer Q-table (PQ) of the query source. For choosing the target peers, another table is constructed by merging the Q-values in NQ and PQ in the descending order of Q-values. Top K nodes (walkers) from the newly constructed table are selected and query messages are routed to them. The benefit of this selection process is the availability of well performing nodes for the successful completion of the query. The techniques for message identification, routing of duplicate messages, TTL enhancement and query load balancing are similar to the original proposal in [1]. TTL enhancement is utilised by power peers to continue searching beyond the given TTL limit.

## III. Q-TABLE UPDATE

This section explains how the Q-learning process is employed by neighbours, and power peers to compute rewards and update Q-values in different Q-tables. When a requested object is found, all the peers on the reverse path update the Q-values. Three kinds of changes may occur during Q-table update: Q-value increase, Q-value decrease, or no change in Q-value. A few attributes are utilised for each update process. Using various attribute values, the reward is computed which is followed by the update of Q-values. If a node does not return any result for the query, it leads to reduction of its Q-value in the Q-tables. At the same time, if a node makes positive results, it causes Q-value increment. Q-value does not change if the node is not a candidate for searching. Due to the frequent changes in Q-values of peers, initially, the nodes being chosen for one action may not be the candidate for another search operation.

***Query Q-table Update:*** The query Q-table maintains Q-values corresponding to each query keyword for all the neighbours. The neighbours are ranked for each keyword in line with the Q-values they possess. Hence, the query Q-table points to the best possible nodes that may host the concerned file containing the query keyword. The search results influence the alteration of Q-values in the query Q-table. The attributes, which are inured to modify the Q-values, are number of hops being visited and the number of search results returned. If a search process takes less number of hops for finding out the required object irrespective of the allocated TTL value, the search efficiency is improved further. At the same time if more number of results are returned, it means that the node is hosting quite a few objects of same subject area and such nodes belong to one category of nodes called specialized nodes. The Q-value update in query Q-table involves calculation of rewards founded on the above-mentioned parameters and applying the rewards to modify the current Q-value.

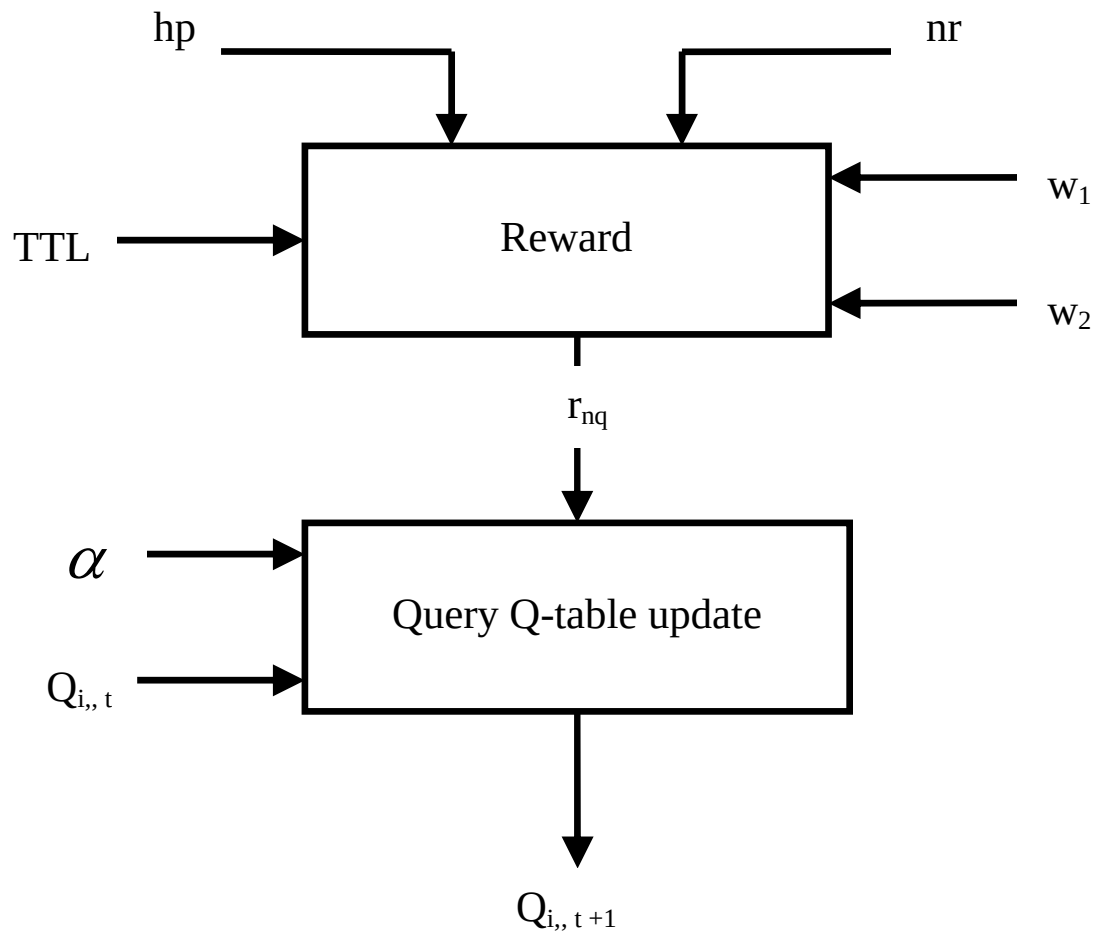

**Figure 1. Query Q-table update**

For each search hit, reward is computed and Q-values in Query Q-tables of intermediate nodes between requester node and node, in which the object is found, are updated on the reverse path. Each result carries a reinforcement signal containing the number of hops (hp) being visited by the peer and the number of results (nr) returned for the query. The reinforcement signal is converted into a reward function as,

$$r_{nq} = \left( \left[ \frac{TTL}{w_1 * hp} \right] + \left[ \frac{nr}{w_2} \right] \right) * 100$$

where $w_2 > w_1$ and the values of $w_1$ and $w_2$ are in between 0 and 1, $w_1 + w_2 = 1$. Specialized nodes may generate a number of matching results; for this reason, weight $w_2$ is made higher than $w_1$. The Q-values in the Query Q-table is updated for a particular query word using the Q-function $Q_{i,t+1} \leftarrow Q_{i,t} + \alpha(r_{nq} - Q_{i,t})$, where α is the learning rate. The Q-values of neighbours (walkers) that positively responded are only updated in this way. All other walkers receive a negative reinforcement $(r_{nq} = 0)$. The reward of those nodes are zero, the Q-value is updated as $Q_{i,t+1} \leftarrow Q_{i,t}(1-\alpha)$. The neighbours who have not participated in the search process keep the Q-values as such, i.e. $Q_{i,t+1} \leftarrow Q_{i,t}$.

If the query keyword is one, which does not exist in the query Q-table, search is further extended through neighbours and power peers. In addition to that, the same keyword is inserted into the query Q-table and the corresponding entries for all the neighbours in the table are initialised with the present Q-values of the neighbour Q-table.

In case the required object is found through a neighbour in the neighbour Q-table, Q-values are modified for the keyword in query Q-table based on number of results returned and number of hops visited. The new entry is deleted from the query Q-table, if the object is not found through the neighbours in the neighbour Q-table.

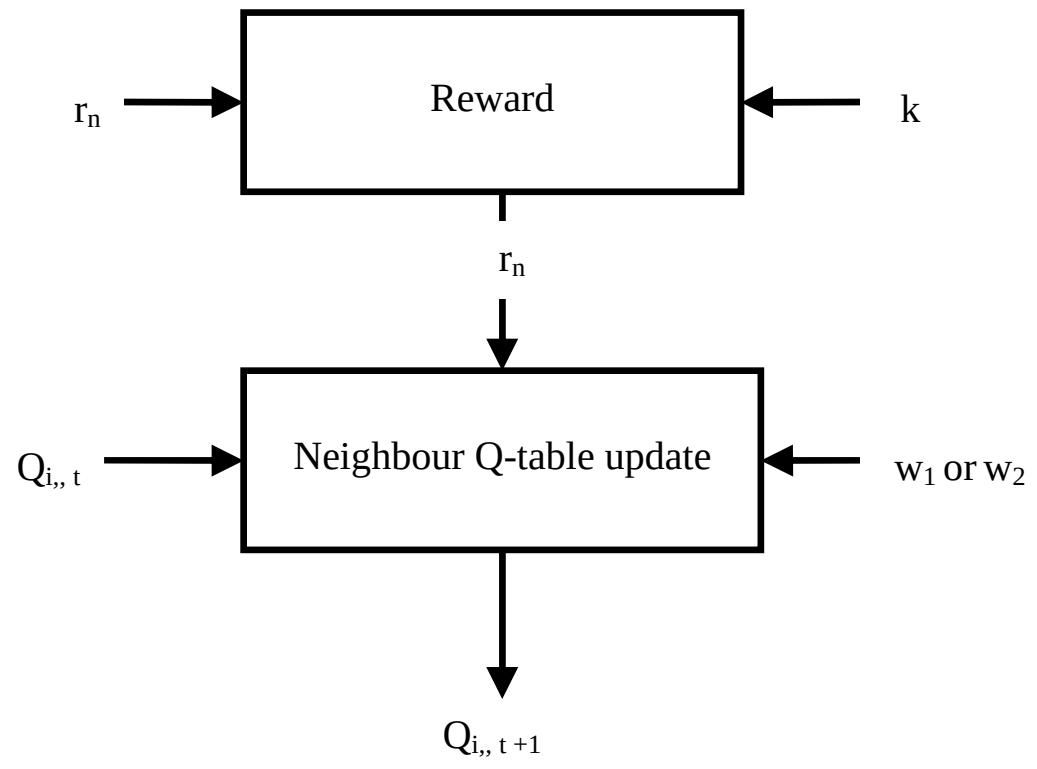

**Figure 2. Neighbour Q-table update**

***Neighbour Q-table Update:*** The query Q-table is concerned only with the performance of neighbours of a node for each query keyword. The overall performance of a neighbour in terms of all the search queries processed by a node is not addressed in a query Q-table. Each node maintains a table called neighbour Q-table to track the performance of all the neighbours of a node. The neighbour Q-table is updated after each query search operation all the way through neighbours. A search hit alters the hit rate of neighbouring nodes. For this reason, the reward for a search operation is computed as $r_n = k * h_r$ where 'k' is a normalizing constant and it is preset as 100 and '$h_r$' is the hit rate. Since the reward $r_n$ is associated with a single variable; the Q-value can be worked out directly. This also reduces the number of operations during update. Moreover, an ordinary peer should be more competent against power peers for producing more hits. Hence, the hit-producing neighbours should maintain high Q-values. For each query hit, the corresponding Q-value of a node (walker) through which the query is successfully processed is modified as $Q_{i,t+1} \leftarrow Q_{i,t} + \left[\frac{r_n}{w_1}\right]$. In the case the object is not found the Q-value is updated as $Q_{i,t+1} \leftarrow Q_{i,t} - \left[\frac{r_n}{w_2}\right]$. The values of the variables $w_1$ and $w_2$ are in between 0.1 and 1, and $w_1 > w_2$. Q-values of remaining neighbours who have not participated in the searching remain unchanged. The propagation of queries through neighbours and power peers together produce hits at a time from neighbouring nodes, power peers or both. If the hit is through a neighbour, an entry for the query keyword is entered in the query Q-table and the Q-values of all the neighbours for the keyword are initialised with the average of Q-values of all the processed queries in each column of the table. Based on the search results, the Q-values are updated using necessary attribute values by following query Q-table update process.

***Power Peer Q-table Update*:** If past successful search data for a query keyword is not available in the query Q-table, walkers are also selected from power peer list of a node. Q-values are updated if a hit occurs through power peer (walker). The power peers support the TTL enhancement feature for extending the search process through other power peers to find out the desired object. Thus, the number of hops visited in a search operation is increased further. Though TTL improvement creates overhead in the network, the advantage is that the success rate is improved. TTL enhancement is necessary in situations where the objects are not properly replicated to suitable sites.

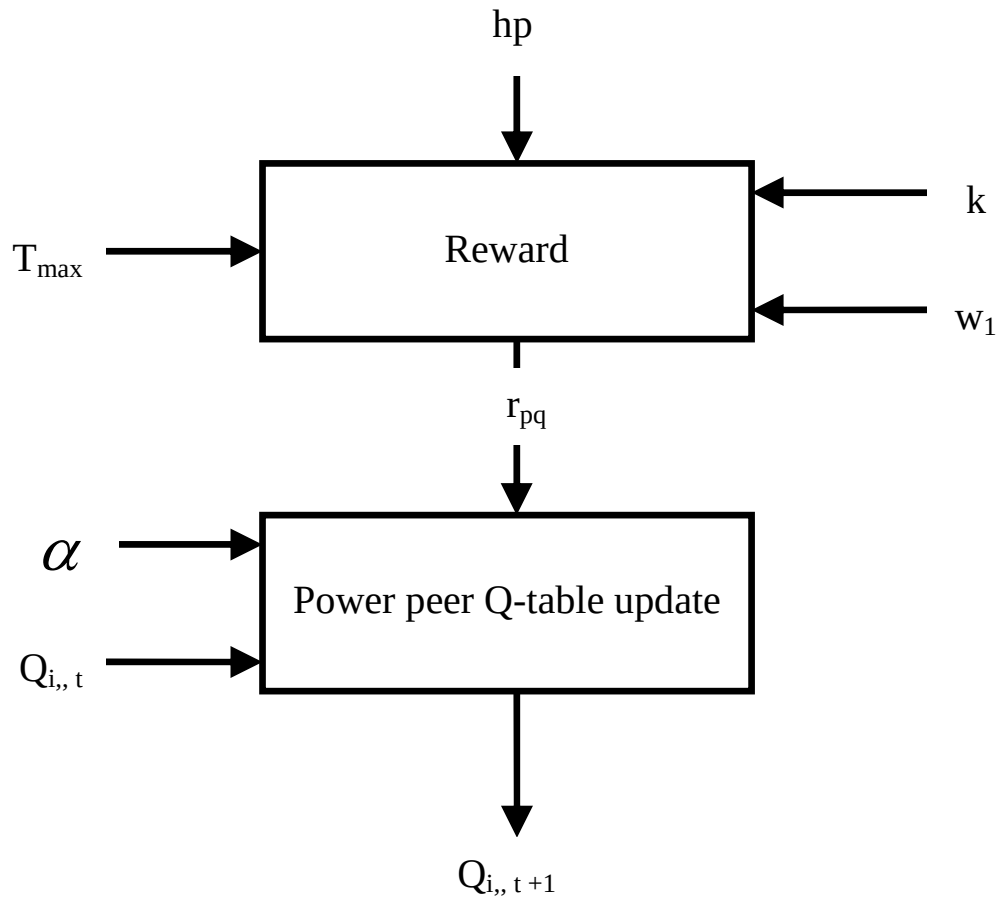

**Figure 3. Power peer Q-table update**

The update of Q-values corresponding to power peers in a power peer Q-table utilises the number of hops visited by each search operation and the maximum permitted TTL value. Due to TTL enhancement, the maximum TTL allowed becomes the actual TTL plus half of the present TTL value. The reward ($r_{pq}$) for each search process is computed based on maximum permitted TTL ($T_{max}$) and number of hops visited (hp).

$$T_{max} = TTL + round\left[\frac{TTL}{2}\right],$$

$$r_{pq} = \left[\frac{T_{max}}{hp * w_1}\right] * k$$

, where K =100 for normalizing the values and the value of $w_1$ is greater than zero and less than 1.

If there is a hit through a power peer, corresponding Q-value of the power peer is updated as $Q_{i,t+1} \leftarrow Q_{i,t} + \alpha(r_{pq} - Q_{i,t})$. The remaining walkers (power peers) who have not produced a hit, update the Q-values as $Q_{i,t+1} \leftarrow Q_{i,t}(1-\alpha)$. No power peers not participated in the search alters their Q-values for the query. If a search operation takes a few hops to find out the result for a query, the Q-value of the peer is increased or else the update causes a decrease in the Q-values.

## IV. Applying our Approach

The search process in DST is illustrated using a simple example. Figure 4 shows a simple P2P network with six ordinary nodes and three power peers. The neighbours of a node are connected together. Each node maintains three types of Q-tables as shown in Figure 5. The Q-tables depict the status of the peers at the time of inputting a new search query. The number of walkers is limited to 1 and the TTL is set to 3.

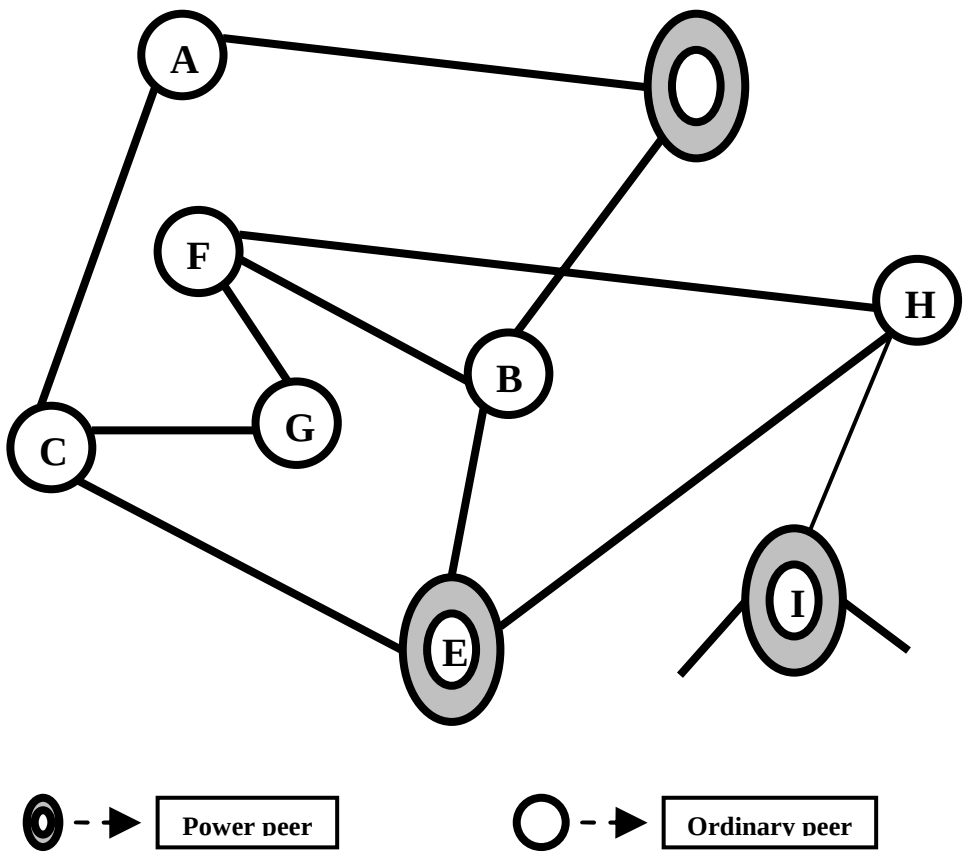

**Figure 4. A Simple P2P Network**

Node 'A' receives a query containing the keyword 'baby' and it checks the shared folder for an object containing the keyword. As the object is not available, the node checks its query Q-table. Since the keyword is present, walker message is generated and it sent to one of the neighbours of node A. Since the neighbour C has the highest Q-value, it receives the message and checks its shared folder for the desired object. The object does not exist in the shared folder of C. In addition, the query Q-table of C does not contain the keyword. As a result, the search is continued to next hop. The node selection policy is applied among the nodes in neighbour Q-table and power peer Q-table of node C. The largest Q-value is held by the neighbour E and the query is routed to it. The desired object is not available in the shared folder of E. While a power peer routes a message through another power peer only, the power peer E selects the target peer for processing the query from its power peer Q-table. The query is routed to another peer 'I', which has the highest Q-value in the power peer Q-table. The required object is available in node 'I' and the walker is terminated. Besides, the TTL value is now reached zero as three hops are already visited.

The node A downloads the required file into its shared folder. The nodes in the search path $\left(A \rightarrow C \rightarrow E \rightarrow I\right)$ update various Q-tables in the reverse direction (Figure 6). The status of Q-tables after the update operations is shown in figure 7. The power peer Q-table of node E, the query Q-table and neighbour Q-table of node C and the query Q-table of node 'A' are updated after the search operation. In the corresponding Q-tables, the Q-values are incremented based on the reward computed. The query word is added to the query Q-table of node C with updated Q-values.

| Node | Neighbor Q-table | Query Q-table | Power Peer Q-table |
|---|---|---|---|
| Node A | C: 346, D: 561 | (C, D)<br>flower: 654, 432<br>baby: 422, 231<br>jack: 653, 411<br>Jill: 234, 476 | E: 567 |
| Node B | D: 543, E: 439, F: 213 | (D, F, E)<br>bag: 321, 765, 542<br>RAM: 432, 178, 333<br>floppy: 187, 233, 342<br>phone: 567, 466, 288 | I: 616 |
| Node C | A: 547, E: 765, G: 632 | (A, E, G)<br>printer: 333, 221, 421<br>book: 653, 432, 337<br>CPU: 568, 164, 421<br>song: 421, 221, 145 | D: 523 |
| Node D | A: 654, B: 556 | (A, B)<br>baby: 456, 344<br>map: 588, 478<br>radio: 290, 304<br>plasma: 314, 378 | E: 412 |
| Node E | B: 688, C: 564, H: 786 | (B, C, H)<br>LCD: 455, 132, 235<br>HDD: 567, 354, 189<br>bus: 764, 267, 521 | D: 546, I: 621 |
| Node F | B: 345, G: 451, H: 254 | (B, G, H)<br>jimmy: 422, 398, 387<br>agent: 356, 289, 189<br>mobile: 378, 299, 192 | D: 543 |
| Node G | F: 548, C: 698 | (F, C)<br>Patrick: 541, 633<br>Harry: 231, 387<br>Potter: 241, 377<br>Johnson: 543, 432 | E: 669 |
| Node H | E: 549, F: 621 | (E, F, I)<br>irrigation: 278, 172, 376<br>snap: 445, 287, 499<br>image: 422, 354, 398<br>map: 266, 511, 546 | D: 515 |

**Figure 5. Status of Q-tables before a new query is submitted**

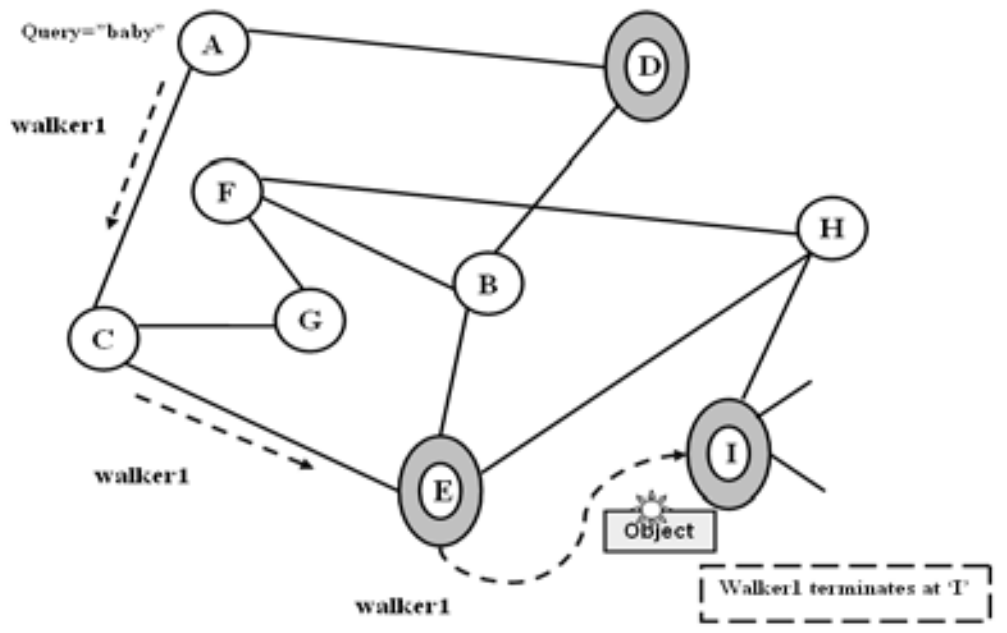

**Figure 6. The Walker Path for a Search Query 'baby'**

| Node | Neighbor Q-table | Query Q-table | Power Peer Q-table |
|---|---|---|---|
| Node A | C: 346, D: 561 | (C, D)<br>flower: 654, 432<br>baby: **669**, 231<br>jack: 653, 411<br>Jill: 234, 476 | E: 567 |
| Node B | D: 543, E: 439, F: 213 | (D, F, E)<br>bag: 321, 765, 542<br>RAM: 432, 178, 333<br>floppy: 187, 233, 342<br>phone: 567, 466, 288 | I: 616 |
| Node C | A: 547, E: **1015**, G: 632 | (A, E, G)<br>printer: 333, 221, 421<br>book: 653, 432, 337<br>CPU: 568, 164, 421<br>song: 421, 221, 145<br>**baby: 547, 956, 632** | D: 523 |
| Node D | A: 654, B: 556 | (A, B)<br>baby: 456, 344<br>map: 588, 478<br>radio: 290, 304<br>plasma: 314, 378 | E: 412 |
| Node E | B: 688, C: 564, H: 786 | (B, C, H)<br>LCD: 455, 132, 235<br>HDD: 567, 354, 189<br>bus: 764, 267, 521 | D: 546, I: **1748** |
| Node F | B: 345, G: 451, H: 254 | (B, G, H)<br>jimmy: 422, 398, 387<br>agent: 356, 289, 189<br>mobile: 378, 299, 192 | D: 543 |
| Node G | F: 548, C: 698 | (F, C)<br>Patrick: 541, 633<br>Harry: 231, 387<br>Potter: 241, 377<br>Johnson: 543, 432 | E: 669 |
| Node H | E: 549, F: 621 | (E, F, I)<br>irrigation: 278, 172, 376<br>snap: 445, 287, 499<br>image: 422, 354, 398<br>map: 266, 511, 546 | D: 515 |

**Figure 7. Status of Q-table after the search operation for query 'baby'**

Consider a scenario with another query keyword 'hello'. The object is not available in node A and its query Q-table. A suitable target node for the walker message is selected from a list prepared by merging the Q-values in neighbour Q-table and power peer Q-table. The power peer 'E' has the highest Q-value and the query message is routed to it and the TTL value is decremented by 1. Since object does not exist in node E, it routes the message to another power peer in its power peer Q-table which is 'I'. The node 'I' hosts the required object and the walker is terminated. Node A downloads the objects from 'I'. The nodes on the search path $\left(A \rightarrow E \rightarrow I\right)$ update the Q-values in different Q-tables in the reverse direction. This is illustrated in figure 8. Figure 9 depicts the status of various Q-tables after the update process. The power peer Q-tables of node E and node A are modified based on the rewards computed.

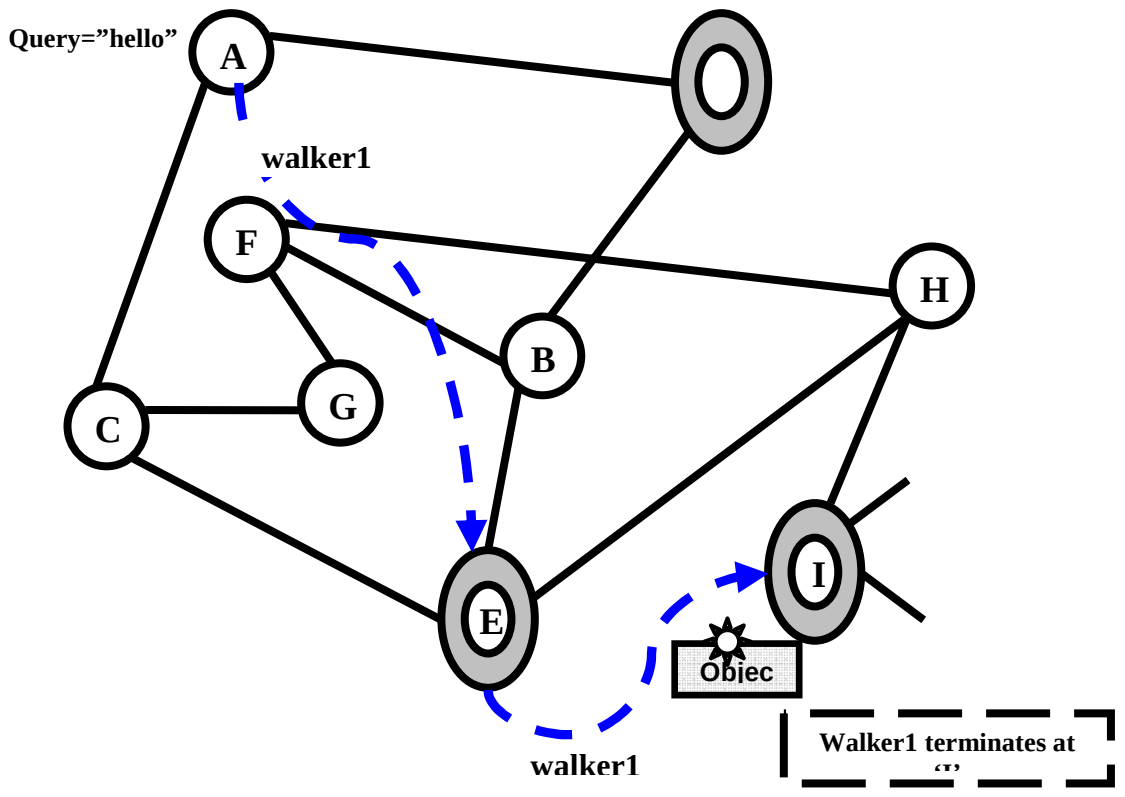

**Figure 8. Search path for query "hello"**

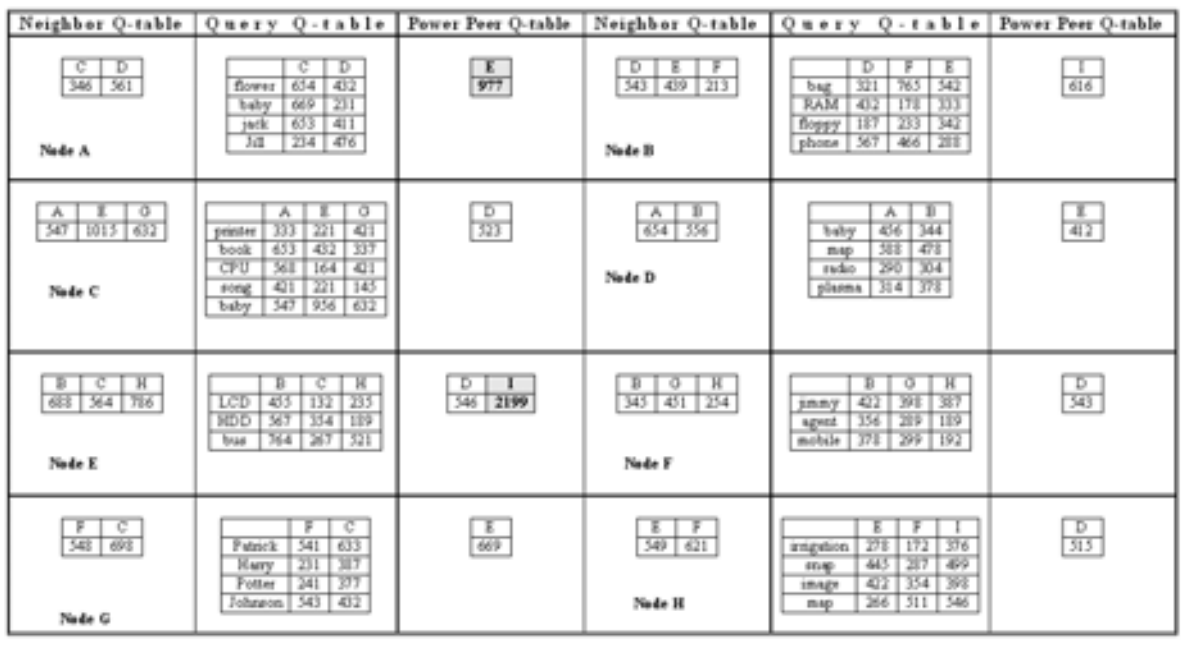

**Figure 9. Q-table Status after the search operation for 'hello'**

## V. EXPERIMENTAL SETUP

The proposed search and replication algorithms are simulated using random graphs that have 10000 nodes. The nodes can join the network and establish random connections to existing nodes. Each node carries a number of files that is greater than or equal to zero, in order to include free riders in the experiments. There are 50 free riders in the network, which do not host any of the files. Each node employs the neighbour selection policy for choosing the neighbours; the free riders are excluded from the neighbour list. Hence, no free riders receive query messages from other nodes. The average degree of a node in the network is 3.5. The value of the learning rate constant is preset as 0.2. There are 100 objects distributed to various nodes, the objects are replicated to different sites using autonomous replication algorithm proposed in [5, 6]. The replication relies on the popularity of the objects. The quantities of objects maintained in the system are sufficient to analyse the performance since autonomous replication scheme effectively propagates the objects to various sites. The objects are word, PDF and text files available as course materials on various subjects such as computer science, electronics, physics, mechanics, electrical etc. Thirty thousand keywords are chosen from the course material files and these words are randomly selected as query keywords by all the nodes during searching. Two types of query search are employed: file name based and keyword based. In the file name based search only the objects names in the shared storage space of each node is searched. The objects containing the keywords are looked up in the keyword-based search. In the simulation scenario, all the queries contain keywords alone.

Each node generates 100 queries and one query is propagated every 20 seconds on average. However, each node enters the query generation phase in a randomly selected time slot. Hence, the flood of query message production is regulated. The least recently used query keywords are removed from the query Q-table to accommodate new keywords. Eighty percent of the nodes are up at the time of performing simulation. Fifty percent of 'Down' nodes selected randomly change their status to 'UP' after every 50,000 queries are propagated and, at the same time, the same amounts of UP nodes obtain the DOWN status. The Q-values in neighbour Q-table and power peer Q-table of each node are initialised with the proposed procedures for initialisation.

The default TTL value is preset as six. There are ordinary nodes and power nodes. The power peers are selected based on node degree, number of objects being hosted and available storage. The minimum degree of a power peer is preset as seven. Initially ten percent of the nodes in the network are assigned the power peer status. The query processing capability of each power peer in terms of CPU_load and free_memory is chosen randomly. The maximum cpu_load that can be processed is represented as number of messages a node can handle at a time. This value remains the same for each power peer until the end of simulation. The availability of free memory relies on the size and quantity of messages. If the total number of messages in the queue reaches the 60% of permitted number of messages in the queue of a power peer, the load data is collected from the power peers listed in the power peer Q-table of the peer according to the load-balancing scheme.

The simulation tool is developed in Java. The tool runs in a Windows operating system environment. The software, which are used for developing the simulation software are NetBeans, J2SE Development Kit 5.0 and WampServer. NetBeans is a free, open-source Integrated Development Environment, which supports development of all Java application types. WampServer is an open source project and Windows web development environment. It allows creating various applications with Apache, PHP and the MySQL database. WampServer also includes PHPMyAdmin and SQLiteManager for managing databases. A mobile agent platform - IBM Aglets can be provided to fire mobile agents to the chosen power peers to collect load data. An aglet can be dispatched to any remote host that supports the Java Virtual Machine. This requires from the remote host to pre-install Tahiti, a tiny aglet server program implemented in Java and provided by the Aglet Framework. To allow aglets (mobile agents) to be fired from within applets, the IBM Aglet team provided the so-called 'FijiApplet', an abstract applet class that is part of a Java package called 'Fiji Kit'. From within the FijiApplet context, aglets can be created, dispatched from and retracted back to the FijiApplet. The simulations are conducted in systems with Intel Xeon (Quad Core) processor,12 MB L2 Cache, 1333 MHz FSB, 4 GB, and 146GB SAS HDD(15K RPM). All the necessary software mentioned above are installed in the systems for conducting various simulation experiments.

Several metrics are employed to evaluate the performance of DST. The performance of the DST algorithm in terms of success rate, average number of hops visited, hits per query, and effect of duplicate messages is compared to two walker-based algorithms: K-random walk [3] and Adaptive Probabilistic Search (APS) [4]. The same table update procedure discussed in [4] is employed for simulating the APS algorithm. The load-balancing performance among power peers is also studied. The impact of TTL enhancement process on successful queries is examined. The role of query cache in query Q-table and simultaneous routing through neighbouring peers and power peers on successful queries are inspected. The advantage of TTL enhancement feature is evaluated. The free riding and partial coverage problems are discussed.

## VI. Results and Discussion

The results of simulation experiments conducted for distributed search technique are explained in this section.

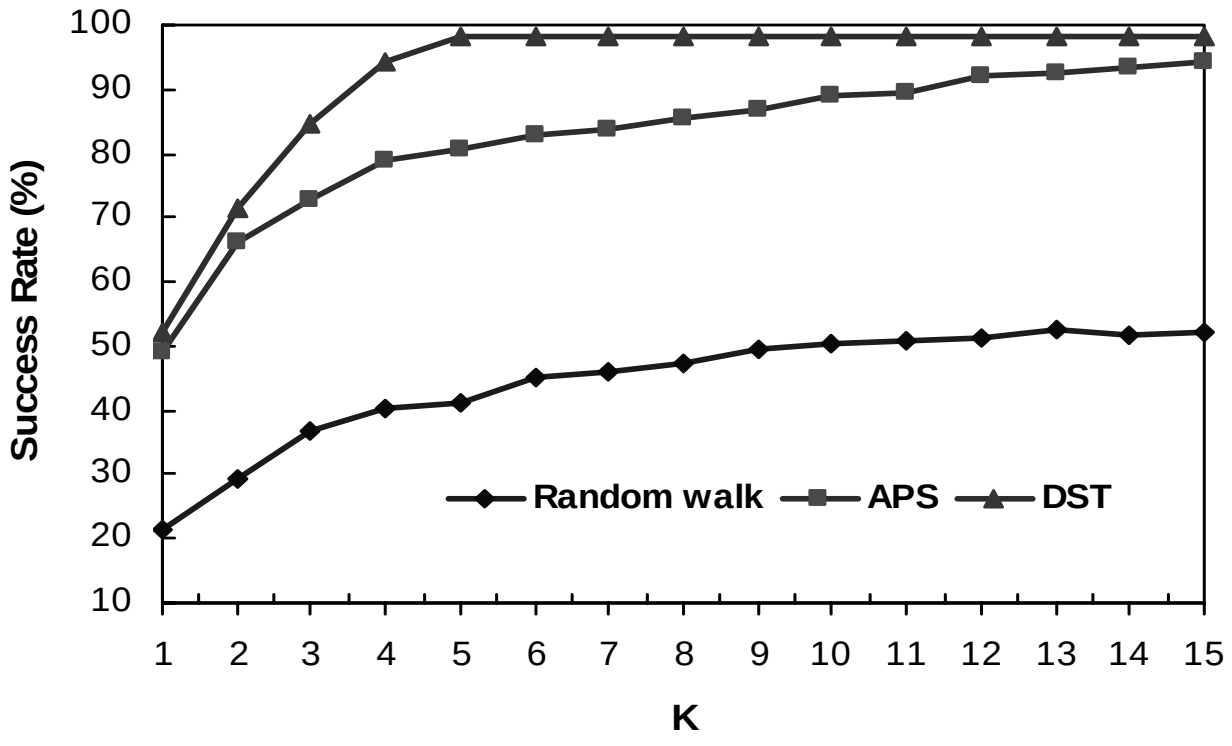

**Figure 10. Comparison of Success Rate**

***Success rate (hit rate):*** The performance of DST in terms of success rate is compared with APS and k-random walk search methods (Figure 10). The experiments are conducted to assess the success rate with different number of walkers (K). The Y-axis of the graph represents the success rate. The performance of DST is superior to APS and random walk for different number of walkers. DST produces high hit rate for very limited number of walkers and TTL value. It achieves more than 98% percentage of hit rate with just six walkers. DST shows a steady success rate for higher value walkers (k>6). However, APS and random walk schemes could not achieve this much of performance even if the number of walkers is increased to 15. The success rate of DST is about 47% greater than random walks and about 17% higher than adaptive probabilistic search with merely six walkers.

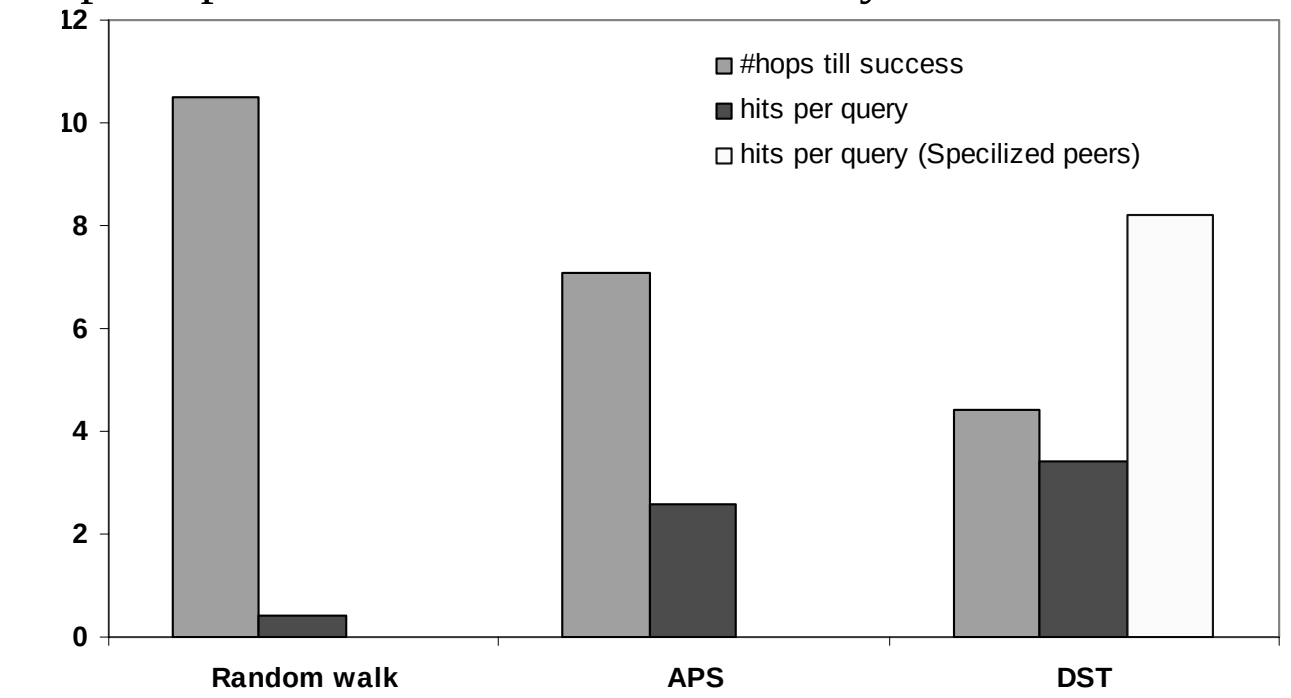

**Figure 11. Average number of Hops visited by a query**

***Number of hops till success:*** DST utilises less number of hops for finding out the objects. This is due to the availability of objects in nearby nodes and the efficient learning mechanism adopted for choosing the target nodes. The average hop distance being used by DST is approximately four for a query, which is less than the number for random walk and APS. APS takes three more hops to complete a query while about six more hops are required for a query in random walk to finish. APS probabilistically picks the target nodes, while random walk selects nodes randomly. However, DST utilises the experience of nodes in routing queries based on Q-learning. Hence, accuracy in routing is enhanced for distributed search technique. At the same time, less number of hops per query saves significant amount network bandwidth. Thus, DST produces high success rate with less cost for processing queries. The average number of hits produced by DST is also shown. DST generates more hits as compared to APS for a few hop counts while the performance of random walk is very low. In DST, as compared to other peers, routing queries through the specialized peers generate more hits depending on the connection between query keywords and objects in the specialized nodes. This is evident from Figure 11 in which DST produces more number of hits per query on specialized nodes.

***Successful queries***: The contribution of both ordinary peers and power peers towards the query success is plotted separately in Figure 12. The contribution of ordinary peers is initially low and it keeps on increasing when more number of queries are processed. The same is applicable to power peers since the performance increases with time. A few factors influence the growth of this success rate. One is that the learning takes time for finding out the exact locations of objects in different nodes in the network. The Q-values of the nodes are modified each time based on the query results. Conversely, the Q-values of better performing nodes are increased. This assists to route the queries through good nodes. When more queries are processed, the popularities of the objects are increased, and the Q-replication scheme creates more number of replicas. These objects are created in well-performing peers. The objects are also copied into query source when the search operation is completed and this further increases the availability of the objects. The increase in number of objects in nodes with sufficient links and storage availability support certain nodes to achieve power peer status. Because of this, the number of power peers in the power peer table of each node increases, and thereby the nodes available for simultaneous routing of queries through both neighbouring nodes and power peers are improved.

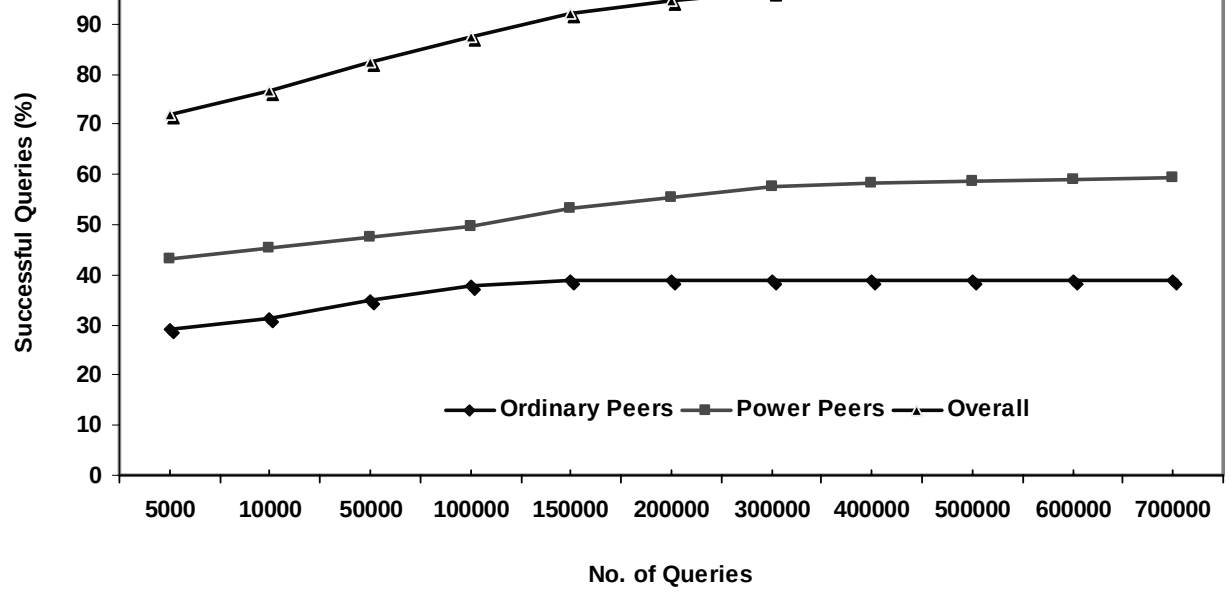

**Figure 12. Contribution of Ordinary peers and Power peers**

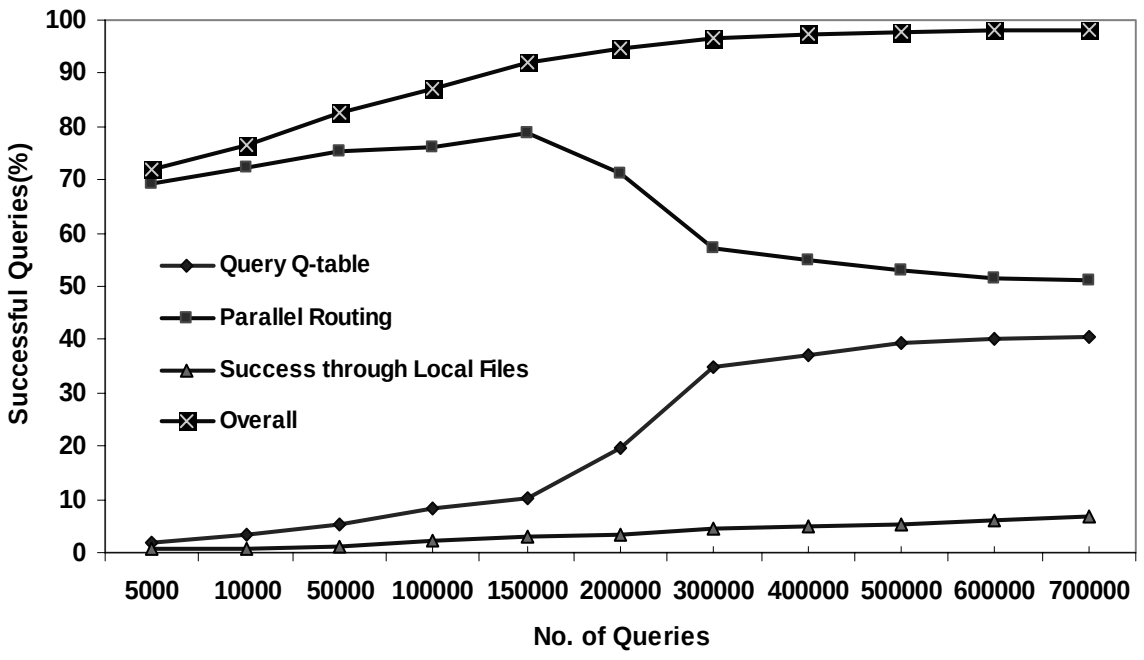

**Figure 13. Query cache, and Parallel Routing**

***Parallel Routing and Query Cache:*** Figure 13 compares the result of routing of queries through nodes in query Q-table and routing of queries in parallel through neighbours in the neighbour Q-table and power peers in the power peer Q-table. The overall success rate relies on the query successes while routing queries through nodes in query Q-table as well as simultaneous routing. Further to that, objects are also directly accessed from the shared storage space of the query source. Initially majority of the objects are discovered through parallel routing. At the start, less number of keywords are available in the query Q-tables; hence, less number of queries are answered through neighbours in the query Q-table. Processing enormous amount of queries increases the size of the query cache in each node and thus the query routing through the neighbours in the query Q-table is augmented. Therefore, the number of successful queries through query Q-table increases with time. Both types of routing achieve a steady state of progress after processing certain amount of queries. This keeps on going until the end of the simulation. This is due to growth of query cache and efficient selection of suitable nodes according to their performance in the past for forwarding the queries. The more popular objects are discovered through nodes in query Q-table, but unpopular objects are found through parallel routing with less number of walkers and small TTL value.

***A load-balancing***: The load on 50 power peers in the network is monitored and data are collected. The performance of load-balancing scheme against no load-balancing is shown in Figure 14. The nodes are arranged in the descending order of load. The proposed load-balancing scheme effectively manages the query load on power peers by distributing the query load among other least loaded power peers.

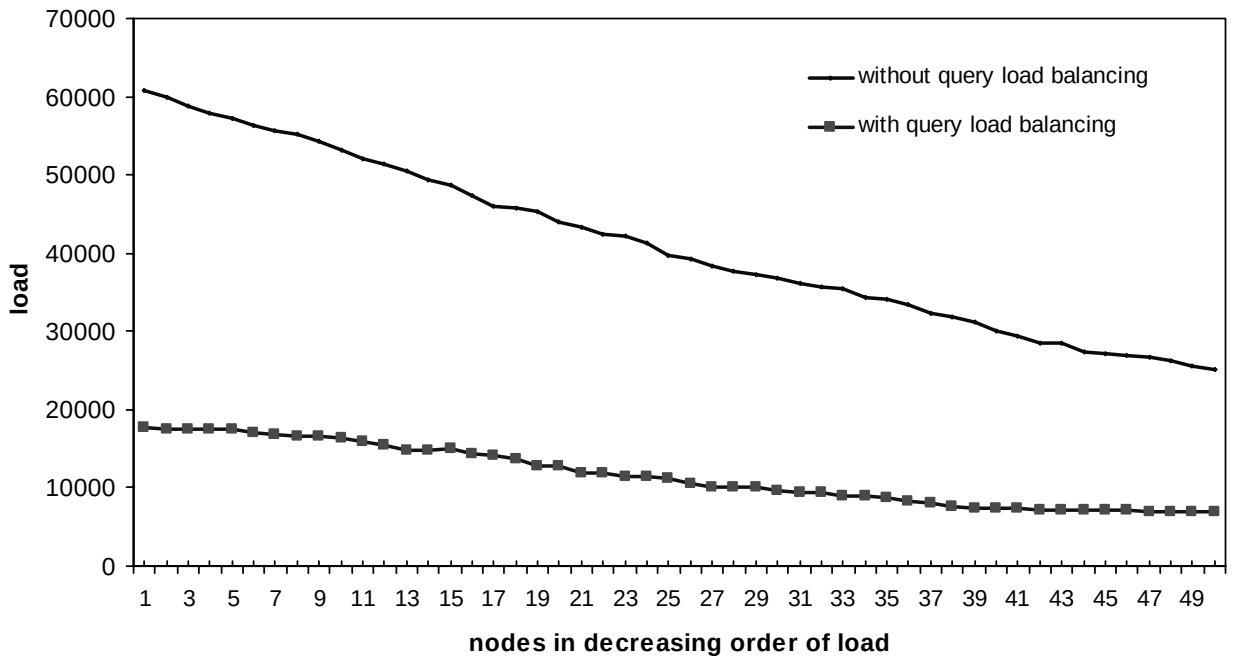

**Figure 14. Query load-balancing**

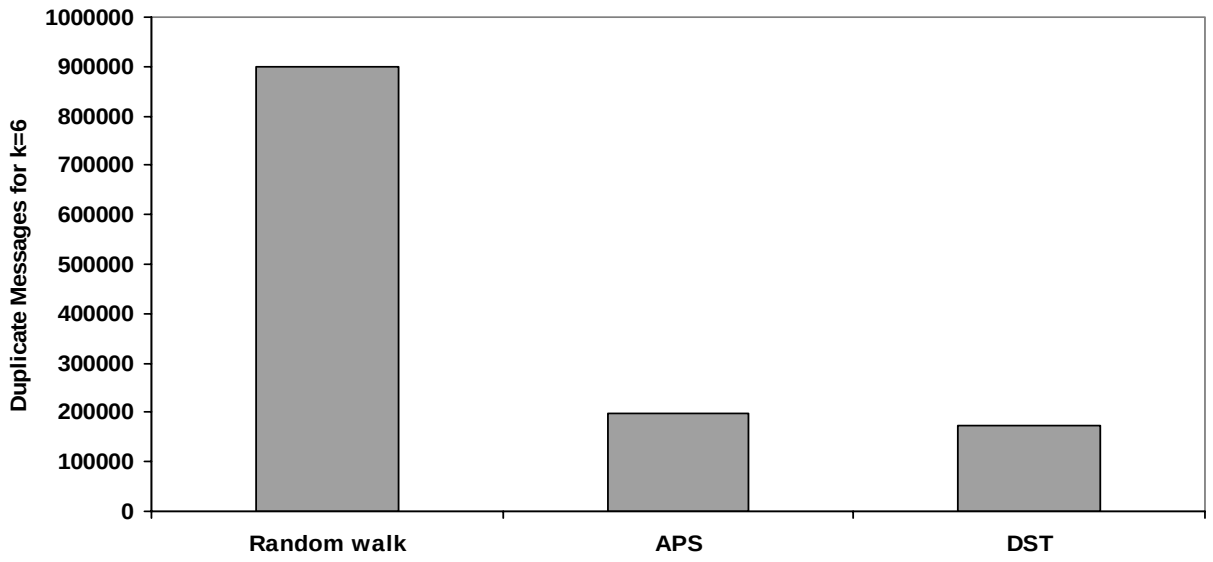

**Figure 15. Comparison of duplicate messages**

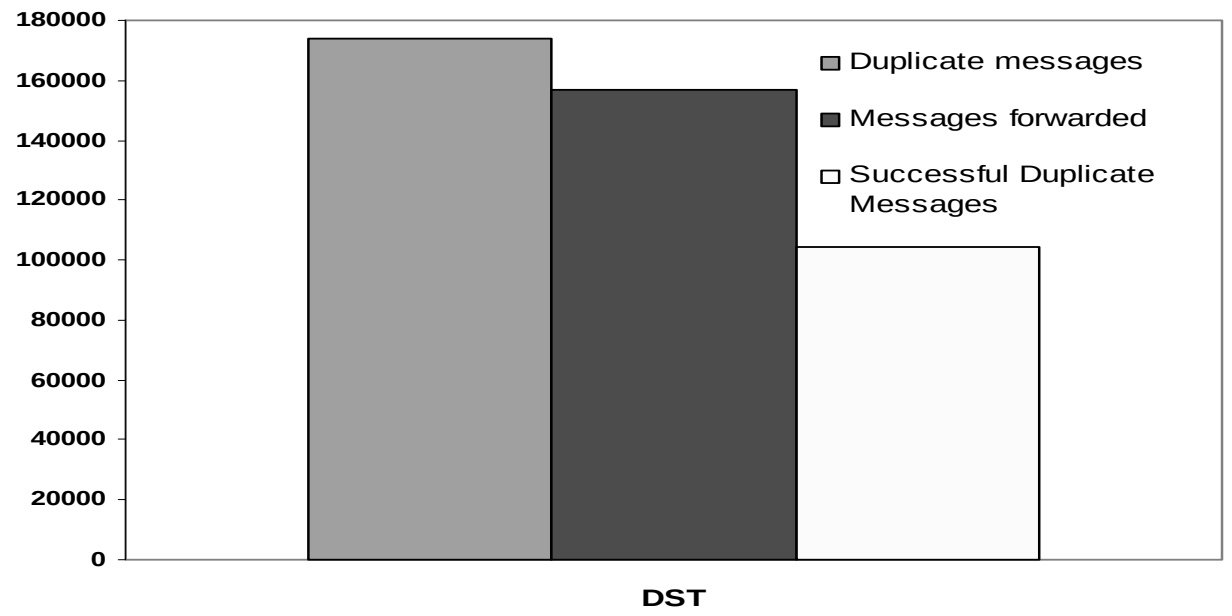

**Figure 16. Utilisation of duplicate messages**

***Duplicate message generated and utilised***: APS considers the duplicate messages as failure states. However, DST routes the messages to other peers subject to the availability of good peers. Random walk produces a much higher number of duplicate messages compared to APS and DST. However, DST generates less number of duplicate messages than APS. The data against six walkers are shown in Figure 15. Figure 16 illustrates that about 90% of the duplicate query messages are effectively routed to other chosen nodes and out of this, 67% of the queries produced hits. About 10% of the duplicate messages could not be forwarded to other nodes, as the available nodes do not meet the criteria for forwarding the duplicate messages.

***TTL Enhancement:*** This feature is employed by power peers to enhance the TTL value for continuing the search operation. The Figure 17 shows how the power peers have utilised the TTL enhancement for increasing the success rate. In the initial stages of simulation, about nineteen percent of the power peers, which are up, have adjusted the TTL values. Gradually in each interval, the number is decreasing due to the availability of objects in nearby nodes and the efficiency of the Q-learning process. At the end of the simulation, less than two percent of the nodes rely on the TTL augmentation process. Similarly, the number of successful queries due to TTL enhancement are decreased from 9.6% to just 1.4% at the end of the simulation (out of total number of successful queries). Hence, the process of TTL augmentation slows down if the popular objects are replicated to good performing nodes and the objects are available in short hop distances.

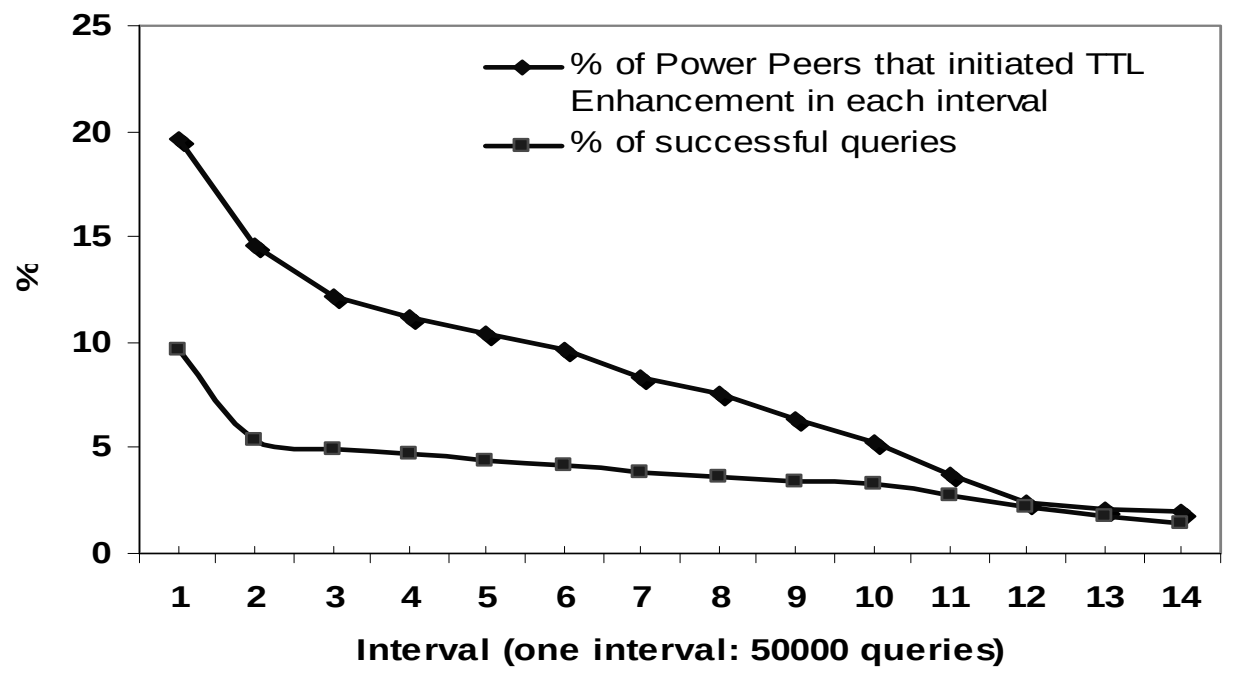

**Figure 17. Performance of TTL Enhancement**

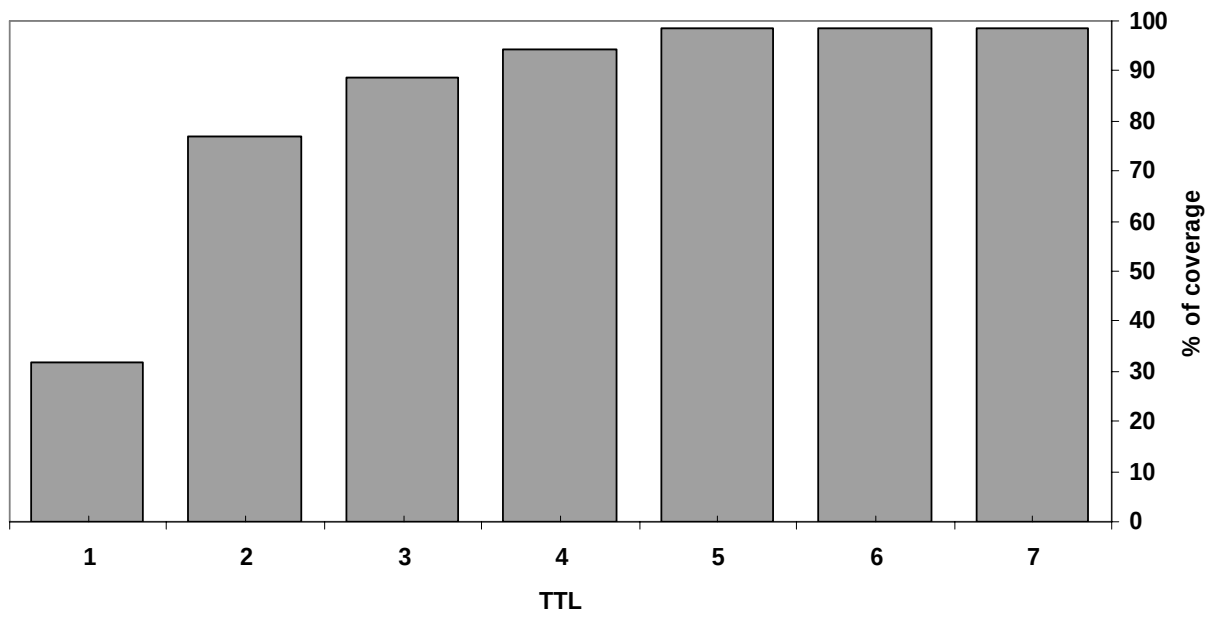

**Figure 18. Partial coverage**

***Partial Coverage and free riding:*** DST successfully alleviates the partial coverage problem by effectively routing queries through best performing nodes. Initially the coverage is low (Figure 18), and this keeps on increasing when more number of queries are propagated. The learning process increases the Q-values of best performing nodes. In each interval, the coverage is growing and it is more than 98% among active nodes for TTL=5 onwards. The query coverage from ordinary nodes and power peers towards free riders is empty. Hence, DST prevents free riders from receiving incoming messages from both the categories of nodes and thus the quantity of unnecessary messages being generated by free riders through the network are reduced. Further to that, the success rate is improved.

## VII. CONCLUSIONS

In this paper, a distributed search algorithm for unstructured P2P network, which employs Q-learning for routing queries and balancing query load is proposed. DST utilises the knowledge that each peer collects about its peers to improve the efficiency of the search. DST employs routing indices to select the target peer for routing queries. The searching does not employ either random or probabilistic forwarding of queries. The load is balanced in two ways: by passing queries simultaneously through ordinary peers and power peers and employing query load-balancing based on Q-learning. Duplicate messages are effectively routed and if the peer is not capable of handling duplicate messages, further incoming duplicate messages are discarded. The various query tables update mechanisms and important data structures related to DST are described. The process of selecting neighbours and performance based searching controls the free riders to some extent. Another advantage of DST is the elimination of the partial coverage problem.